\documentclass[amsmath,amssymb,aps,prx,reprint,superscriptaddress]{revtex4-2}

\usepackage{amsmath}
\usepackage{xcolor}
\usepackage{amsfonts,amssymb}
\usepackage{graphicx}
\usepackage{hyperref}
\usepackage{academicons}
\usepackage{multirow}

\begin{document}

\title{Modeling phase transformations in Mn-rich disordered rocksalt cathodes with machine learning interatomic potentials}

\author{\text{Peichen Zhong}}
\email[]{zhongpc@berkeley.edu}
\affiliation{Department of Materials Science and Engineering, University of California, Berkeley, California 94720, United States}
\affiliation{Materials Sciences Division, Lawrence Berkeley National Laboratory, California 94720, United States}

\author{\text{Bowen Deng}}
\affiliation{Department of Materials Science and Engineering, University of California, Berkeley, California 94720, United States}
\affiliation{Materials Sciences Division, Lawrence Berkeley National Laboratory, California 94720, United States}

\author{\text{Shashwat Anand}}
\affiliation{Materials Sciences Division, Lawrence Berkeley National Laboratory, California 94720, United States}

\author{\text{Tara Mishra}}
\affiliation{Materials Sciences Division, Lawrence Berkeley National Laboratory, California 94720, United States}

\author{\text{Gerbrand Ceder}}
\email[]{gceder@berkeley.edu}
\affiliation{Department of Materials Science and Engineering, University of California, Berkeley, California 94720, United States}
\affiliation{Materials Sciences Division, Lawrence Berkeley National Laboratory, California 94720, United States}

\date{\today}


\begin{abstract}
Mn-rich disordered rocksalt (DRX) cathode materials exhibit a phase transformation from a disordered to a partially disordered spinel-like structure ($\delta$-phase) during electrochemical cycling. In this computational study, we used charge-informed molecular dynamics with a fine-tuned CHGNet foundation potential to investigate the phase transformation in Li$_{x}$Mn$_{0.8}$Ti$_{0.1}$O$_{1.9}$F$_{0.1}$. Our results indicate that transition metal migration occurs and reorders to form the spinel-like ordering in an FCC anion framework. The transformed structure contains a higher concentration of non-transition metal (0-TM) face-sharing channels, which are known to improve Li transport kinetics. Analysis of the Mn valence distribution suggests that the appearance of tetrahedral Mn$^{2+}$ is a consequence of spinel-like ordering, rather than the trigger for cation migration as previously suggested. Calculated equilibrium intercalation voltage profiles demonstrate that the $\delta$-phase, unlike the ordered spinel, exhibits solid-solution signatures at low voltage. A higher Li capacity is obtained than in the DRX phase. This study provides atomic insights into solid-state phase transformation and its relation to experimental electrochemistry, highlighting the potential of machine learning interatomic potentials for understanding complex oxide materials.
\end{abstract} 


\maketitle

\section{Introduction}

The increasing demand for sustainable energy storage has driven the development of high-energy-density and cost-effective cathode materials for rechargeable Li-ion batteries \cite{Olivetti2017, Xie2021_carbon_neutral, Tian2021_promise}. Disordered rocksalt (DRX) cathodes using manganese (Mn) as the earth-abundant redox-active cation have the potential to scale Li-ion energy storage to several TWh/year \cite{Clement2020_DRX_review}. Recent studies demonstrate that Mn-rich disordered rocksalt structures undergo an in-situ transformation into a partially disordered spinel-like phase (termed the $\delta$-phase) upon electrochemical cycling \cite{Kwon2020, Li2021_AFM_TEM, Zhou2021_LMTO, Ahn2023_AEM, Cai2023_NatEnergy}, which results in increased capacity and rate capability for cathode materials.
This approach mitigates the typical low capacity in other DRX cathodes by promoting a new voltage feature around 4 V and an extended 3 V plateau. Moreover, in contrast to conventional Li$_x$Mn$_2$O$_4$ spinel cathodes, which suffer from a two-phase reaction when fully lithiated between $0.5< x< 1$, leading to inhomogeneity-induced stresses and mechanical degradation, the $\delta$-phase maintains a single (solid-solution) phase reaction throughout the cycling process. \citet{Cai2023_NatEnergy} demonstrates that such $\delta$-phase material exhibits an impressive capacity of over 280 mAh/g and an energy density exceeding 800 Wh/kg while maintaining minimal voltage decay even after extensive cycling.

\begin{figure*}[htb]
\centering
\includegraphics[width=0.95\linewidth]{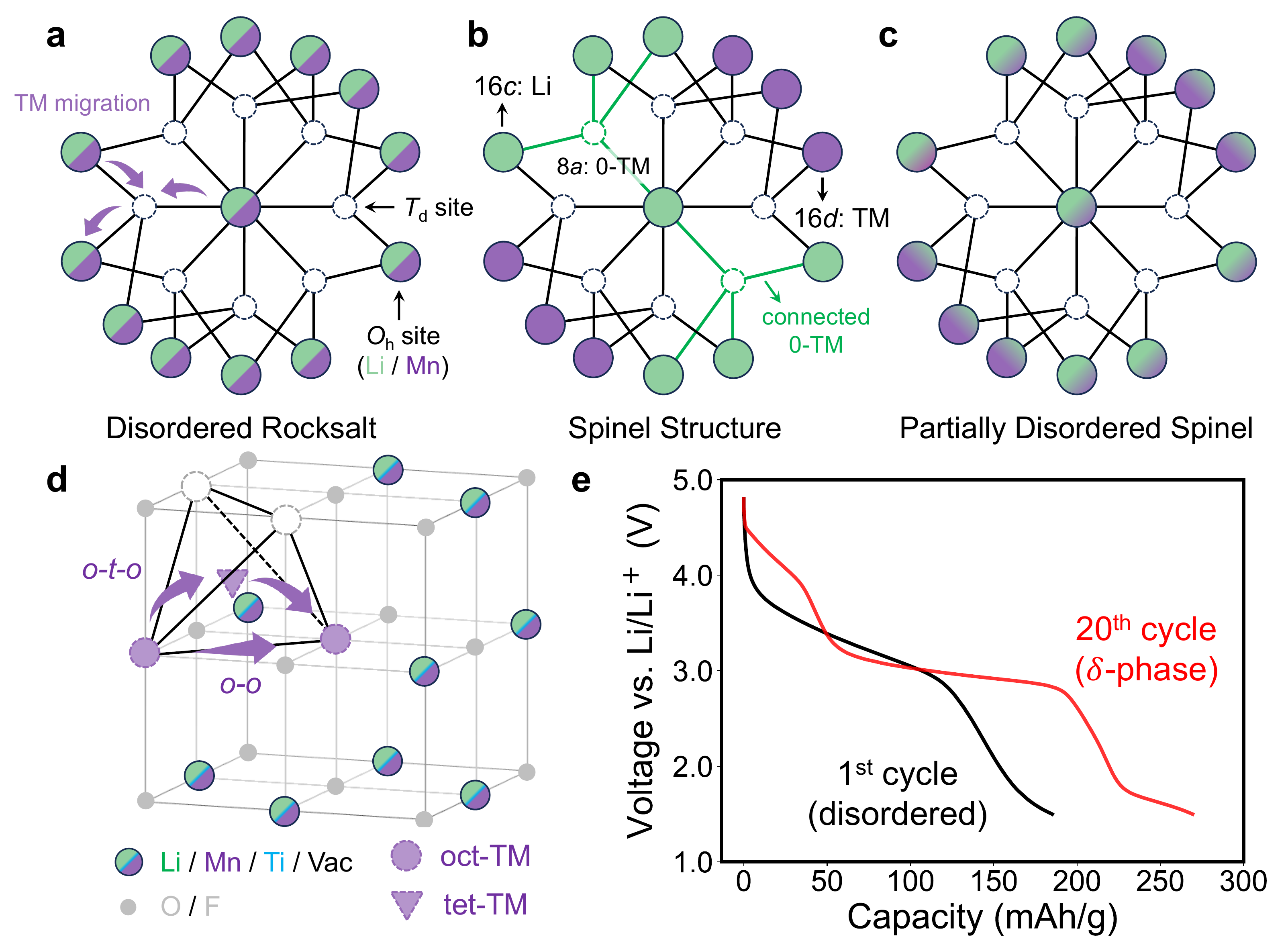}
\caption{\textbf{Local ordering and electrochemical characteristics in Li-Mn-(Ti)-O-F systems}. (a-c) Schematic representations of local cation ordering in LiMnO$_2$: (a) disordered rocksalt, (b) spinel, and (c) partially disordered spinel structures. Green and purple circles represent Li and Mn occupying octahedral ($O_h$, denoted as $o$) sites in an FCC rocksalt framework, respectively. Edges connect $O_h$ sites to interstitial tetrahedral sites ($T_d$, denoted as $t$). In the spinel structure (b), $O_h$ sites are categorized as $16c$ (Li-occupied) and $16d$ (Mn-occupied). The $T_d$ sites adjacent to $16c$ Li are designated as $8a$ sites in 0-TM tetrahedra (without TM neighbors), forming an interconnected percolating network (green lines). The partially disordered spinel structure (c) shows intermediate $16c/16d$ site occupancy between spinel and disordered structures. (d) TM migration pathways in the FCC rocksalt framework: arrows indicate possible migration routes through an intermediate $T_d$ site ($o$-$t$-$o$) or direct connection between $O_h$ sites ($o$-$o$). (e) Experimental voltage profiles of Li$_{x}$Mn$_{0.8}$Ti$_{0.1}$O$_{1.9}$F$_{0.1}$ during cycling between 1.5 and 4.8 V at 20 mA/g from Ref.~\cite{Cai2023_NatEnergy}, showing the 1st (black line) and 20th (red line) cycles. The 20th cycle reveals a new voltage feature at ~4 V and an extended 3 V plateau, indicating $\delta$-phase formation.}
\label{fig:spinel_intro}
\end{figure*}

\begin{figure*}[t]
\centering
\includegraphics[width=\linewidth]{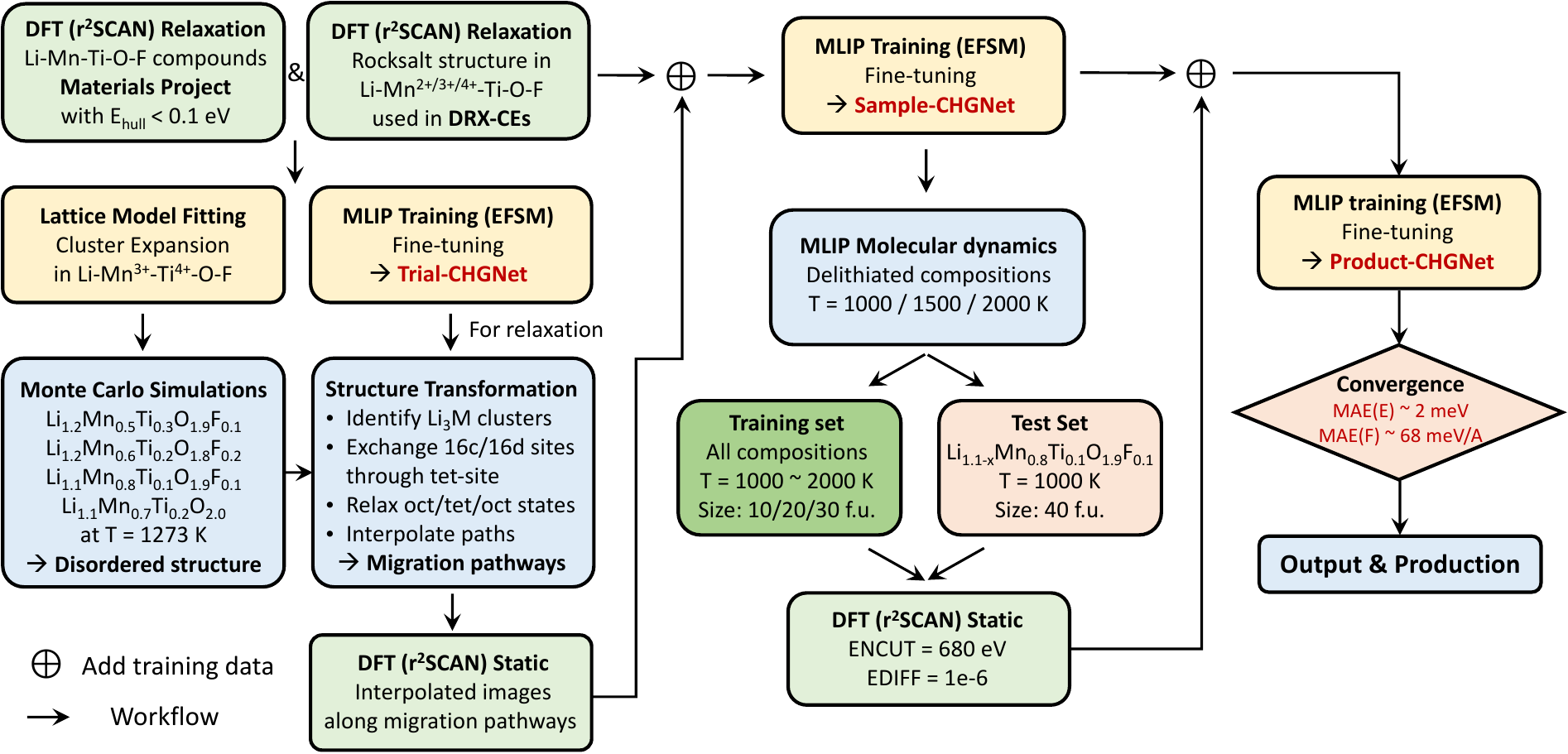}
\caption{Flowchart illustrating the development process of fine-tuned CHGNet machine learning interatomic potential (MLIP) and cluster expansion models for the Li--Mn--Ti--O--F system.}
\label{fig:finetune_LMTOF} 
\end{figure*}

The spinel is distinguished from the disordered rocksalt by its cation ordering in the common FCC anion framework. Figure \ref{fig:spinel_intro}a--\ref{fig:spinel_intro}c presents the cation orderings in a conceptualized graph representation as a projection of the 3-dimensional FCC structure shown in Fig.~\ref{fig:spinel_intro}d, including the disordered rocksalt (Fm$\bar{3}$m), spinel (Fd$\bar{3}$m), and partially disordered spinel ($\delta$-phase) structures. 
The green and purple spheres represent Li and Mn ions, respectively, occupying the octahedral sites ($O_h$, denoted as $o$) of the FCC rocksalt framework.
For simplicity, the stoichiometric LiMnO$_2$ is used for illustration in Fig.~\ref{fig:spinel_intro}a-c. 
The graph edges connect the $O_h$ sites to the interstitial tetrahedral sites ($T_d$, denoted as $t$), which are represented by white dashed circles in Fig.~\ref{fig:spinel_intro}. 
In the spinel LiMnO$_2$ ordering, the octahedral sites are categorized into $16c$ and $16d$ sites as shown in Fig.~\ref{fig:spinel_intro}b. All $16c$ sites are occupied by Li, while all $16d$ sites are occupied by Mn. 
Note that we refer to spinel LiMnO$_2$ here as the product of lithiation of the traditional spinel LiMn$_2$O$_4$.
The tetrahedral site neighbored by four $16c$ Li is denoted as an $8a$ site in a spinel. When the spinel is delithiated to Li$_{0.5}$MnO$_2$, the $16c$ sites become vacant and the $8a$ sites are occupied by Li, representing a traditional spinel structure. More generally, the tetrahedral site neighbored by four Li is also termed a 0-TM channel. \citet{Urban2014_AEM_order} demonstrated that long-range connected 0-TM percolating channels are essential for facile Li diffusion in an FCC framework, with the spinel ordering having the highest amount of 0-TM channels of any ordering. The partially disordered spinel is conceptually illustrated in Fig.~\ref{fig:spinel_intro}c as an intermediate state between the spinel and the disordered rocksalt.

The phase transformation from the DRX to $\delta$-phase is related to the transition metal (TM) migration of the mobile Mn ion when neighboring cation vacancies are created \cite{Reed2001_LiMnO}. The TM migration can occur along the pathway connected by the octahedral-tetrahedral-octahedral ($o$-$t$-$o$) sites in an FCC framework. The Mn migration barrier is strongly coupled with its charge state: Mn$^{4+}$ is generally considered immobile, while Mn$^{3+}$ and Mn$^{2+}$ are considered more mobile according to the nudged-elastic band calculations \cite{Jang_Chou_Huang_Sadoway_Chiang_2003, Jo2019_layer_Mn_migration, radin2019_NE_Mn7}. 
However, simulating these charge-coupled dynamics has long been challenging, as the required time and length scales are inaccessible to conventional \textit{ab initio} methods.
Previously, cluster expansion (CE) methods have successfully acted as a surrogate model to density functional theory (DFT) calculations for modeling the configurational thermodynamics in complex oxides via Monte Carlo (MC) simulation at scale \cite{Squires2023_SRO, Puchala2023_CASM, Zhong2023_PRXEnergy}. Nonetheless, the limited expressibility and coarse-grained nature of these models render them less adequate for capturing the complex charge-coupled dynamical processes involved in TM migrations \cite{Chen2023_remove, Cai2023_NatEnergy}. The advent of machine learning interatomic potentials (MLIPs) has opened new avenues for addressing these problems by enabling large-scale and nano-second-long simulations with \textit{ab-initio} quantum accuracy \cite{Batzner2022_nequip, chen2022_m3gnet, Deng2023_chgnet}. 
By using magnetic moments as a proxy for atomic charge, \citet{Deng2023_chgnet} recently simulated a phase transformation from an orthorhombic to a spinel-like phase in Li$_{0.5}$MnO$_2$ with charge-informed MD.
It identified the timescale discrepancies between charge disproportionation and ion hops, suggesting that the emergence of Mn$^{2+}$ is correlated to long-range spinel-like ordering but not solely to the local hopping into the tetrahedral occupancy.

In this study, we study the complex phase transformation in Li$_{x}$Mn$_{0.8}$Ti$_{0.1}$O$_{1.9}$F$_{0.1}$ (LMTOF), which is a Mn-rich DRX cathode material. We parametrized a CHGNet-MLIP using r$^2$SCAN-DFT calculations for various atomic configurations to sample the potential energy surface in this chemical space. The fine-tuned CHGNet-MLIP was used to implement charge-informed MD simulations starting from a disordered phase in Li$_{0.6}$Mn$_{0.8}$Ti$_{0.1}$O$_{1.9}$F$_{0.1}$. The MD simulations reveal a phase transformation from a disordered rocksalt to a spinel-like ($\delta$) phase. From the magnetic moment predictions, we demonstrate that the Mn migration mainly occurs as Mn$^{3+}$ in the initial part of the transformation, while Mn$^{2+}$ collectively emerges as the spinel-like ordering forms. Furthermore, we investigated the impact of this partial disorder on the electrochemical intercalation voltage profiles using the structures obtained from the MD simulations. Our work provides insights into the application of MLIPs for studying complex energy materials and offers a mechanistic understanding of the partial cation-disorder effect on intercalation electrochemistry at the atomic scale.

\section{Methods}

\subsection{Machine learning interatomic potential}

To efficiently compute the energy and interatomic forces, we employed the Crystal Hamiltonian Graph Neural Network (CHGNet) as a surrogate model for DFT calculations \cite{Deng2023_chgnet}. CHGNet is a graph-neural-network-based MLIP that utilizes message-passing layers to propagate atomic information via a set of nodes $\{v_i\}$ connected by edges $\{e_{ij}\}$ \cite{gilmer2017_MPNN}. After $n-1$ message-passing layers, the node features $\{v_i^{n-1}\}$ are linearly projected onto magnetic moments by $\phi_m(\cdot)$, and the total energy is calculated as a sum of non-linear node projections by $\phi_E(\cdot)$ over all atoms $\{v_i^n\}$ with an additional message-passing layer
\begin{equation}
m_i = \phi_m(v_i^{n-1}), \quad
E_{\text{tot}} = \sum_i \phi_E(v^n_i).
\end{equation}
The nodes $\{v_i^{n-1}\}$ are regularized by DFT magnetic moments to carry rich information about both local ionic environments. The valence state of TM is inferred from the predicted magnetic moment as a proxy of the atomic charge, which is a direct output of the MLIP rather than an input parameter.
The forces and stress are calculated via auto-differentiation of the total energy with respect to the atomic Cartesian coordinates and strain:
\begin{equation}
\vec{f}_i = - \frac{\partial E_{\text{tot}}}{\partial \vec{x}_i}, \quad
\boldsymbol{\sigma} = \frac{1}{V}\frac{\partial E_{\text{tot}}}{\partial \boldsymbol{\varepsilon}}.
\end{equation}
This approach allows CHGNet to capture the complex relationships between charge distributions and energetics with given atomic configurations, which is essential to properly model the kinetic path of evolving TM oxides \cite{Zhou2006_LFP}.

\subsection{Potential energy surface sampling}

CHGNet \cite{Deng2023_chgnet} as a pre-trained foundation potential achieves reasonable performance in predicting materials stability on out-of-distribution datasets \cite{Riebesell_2025}. However, recent benchmarks observed a systematic softening of the ML predicted potential energy surface (PES) compared to the DFT ground truth, particularly for atomic configurations with high energy \cite{Deng_2025, Kaplan_2025}. 
To improve simulation accuracy, we performed multi-faceted sampling of the PES for the Li--Mn--Ti--O--F chemical space \cite{Shyue_JACS}, as follows:

\textit{Exisiting dataset} -- We collected the compounds in the Materials Project database within the Li--Mn--Ti--O--F chemical space that have decomposition energy (energy above the convex hull, $E_{\text{hull}}$) lower than 0.1 eV/atom as a general coverage of such chemical space \cite{Jain2013}. Furthermore, we adopted the structures from the training dataset of previously reported cluster expansion (CE) models for disordered rocksalts containing Mn$^{2+}$/Mn$^{3+}$ as redox-active species (DRX-CE) \cite{Ji2019_hidden_SRO, Ouyang2020_SRO}. All these structures in the DRX-CE dataset have rocksalt lattices but different cation and anion orderings. These structures were relaxed by DFT calculations. The relaxed structures with rocksalt frameworks were used to fit a cluster expansion Hamiltonian (lattice model fitting, see Appendix \ref{appendix:CE}). The relaxation trajectories from structures from the two datasets were used to fine-tune a Trial CHGNet.

\textit{Partially disordered structures} -- 
Given that existing datasets do not include atomic configurations specifically representing partially disordered structures with spinel-like orderings, we developed a custom sampling procedure to generate these configurations. 
(a) Monte Carlo (MC) simulations were employed to obtain disordered rocksalt atomic configurations in a supercell of the primitive spinel cell by equilibration at high temperatures (see Appendix \ref{appendix:CE}). 
(b) The octahedral cation sites are then categorized into $16c$ and $16d$ sites by overlaying the template of the perfectly ordered spinel. Under this classification, a random distribution of TM yields the disordered structure, while preferential TM occupation of $16d$ sites indicates more spinel-like ordering.

To generate different degrees of spinel order, we applied site exchanges between two nearest neighbor octahedral sites to transform the disordered structure towards the spinel ordering through: (a) 
An exchange between a $16c$ TM and a $16d$ Li in a Li$_3$M tetrahedron. We selected a Li$_3$M tetrahedron that contains three Li atoms (including at least one $16d$
Li) and one $16c$ metal atom. A site exchange is applied to the $16d$ Li and the $16c$ TM.
(b) An exchange between a $16d$ TM and a $16d$ Li. 
Step (a) increases the spinel-like order, but it may halt if no suitable Li$_3$M tetrahedra are available to facilitate the exchange. To address this, Step (b) rearranges cations within the $16d$ sites. While this internal exchange does not change the overall degree of spinel ordering, its purpose is to create new Li$_3$M tetrahedra, thereby enabling further transformations via Step (a).
By iteratively applying site exchanges within different Li$_3$M tetrahedra, one can obtain a collection of structures ranging from disordered to spinel ordering.

\begin{figure}[ht]
\centering
\includegraphics[width=\linewidth]{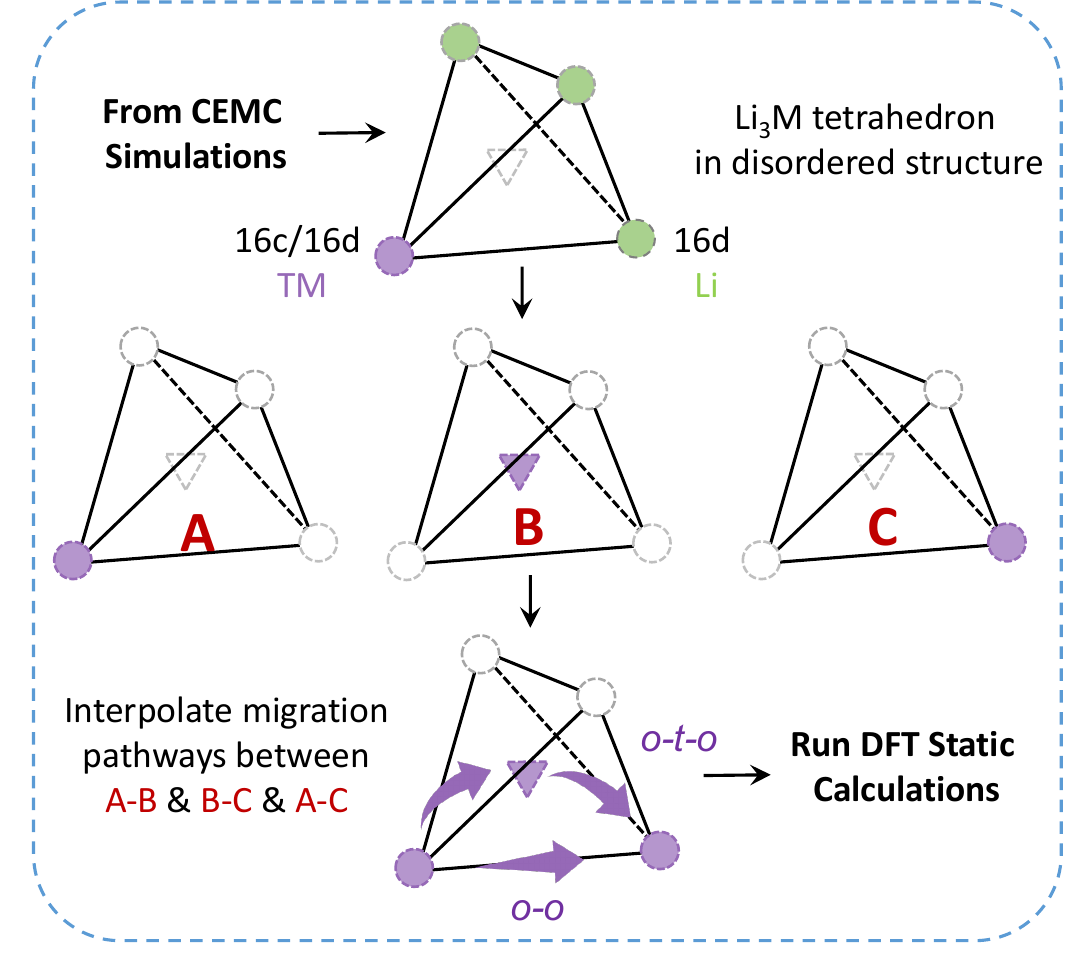}
\caption{\textbf{Schematic representation of TM migration pathway sampling process}. The procedure consists of three steps: (1) Identification of a Li$_3$M tetrahedron containing one $16c$ or $16d$ TM and one $16d$ Li; (2) Creation of vacancies by removing Li ions from $O_h$ sites, followed by TM placement at three positions representing the initial (A), intermediate (B), and final (C) states of $o$-$t$-$o$ migration; (3) Generation of atomic configurations along A-B, B-C, and A-C paths through interpolation, with subsequent evaluation using static DFT calculations.}
\label{fig:local_migration} 
\end{figure}

\textit{TM migration pathways} -- TM migration is a rare, high-energy-barrier event \cite{Cai2023_NatEnergy}, especially for unfavorable valence states such as  Mn$^{4+}$. To effectively model the diffusion kinetics, we implemented a passive sampling approach with the following steps:

A TM can only migrate from its octahedral site into or through a tetrahedral site which has no other face-sharing TMs. In the fully lithiated state, there are Li$_3$M tetrahedra where the tetrahedron is surrounded by only one TM and 3 Li. For each migration event via a Li$_3$M tetrahedron, three different states are generated as A (initial state, TM occupied on $O_h$ site), B (intermediate state, TM occupied on $T_h$ site), and C (final state, TM occupied on $O_h$ site), which are shown in Fig.~\ref{fig:local_migration}. The Li atoms face-sharing with the $T_h$ site are removed to create vacancies so that there are no face-sharing ions with the migrating TM and $O_h$ cations. The structures in the A/B/C states were relaxed using the Trial-CHGNet. After structure relaxation, five images interpolated along A-B ($o$-$t$ path) and B-C ($t$-$o$ path) are evaluated with static DFT calculations, analogous to the intermediate states for nudged elastic band (NEB) calculations. Furthermore, we follow the same approach to generate states along a direct migration between $O_h$ sites, which cuts through their shared edge (A-C) so that high-energy states are considered in the MLIP fitting \cite{VanderVen2001_oto}.

In addition, we considered two different delithiation levels for sampling TM migration pathways:
(a) A dilute level, where only three Li ions are removed to create a diffusion channel for TM migration (e.g., a pristine supercell structure with Li$_{44}$Mn$_{32}$Ti$_{4}$O$_{76}$F$_{4}$ yields a delithiated structure with Li$_{41}$Mn$_{32}$Ti$_{4}$O$_{76}$F$_{4}$). This level allows for the investigation of the environment dependence of the TM hopping energetics.
(b) A medium delithiation level, where 0.5 Li per formula unit (f.u.) is removed from the structure to create delithiated states (e.g., Li ions are randomly removed to create a structure with Li$_{24}$Mn$_{32}$Ti$_{4}$O$_{76}$F$_{4}$). This level allows the sampling of delithiated states, representing the charging process in a battery. By sampling partially disordered structures at these two delithiation levels, we include a more comprehensive description of the cation ordering energetics, TM hop, and relevant local environments in which TM hopping occurs at the composition that is energetically favorable for spinel formation (i.e., the composition close to Li$_{0.5}$TMO$_2$) \cite{Ceder1999_phasediagram}. By adding the static DFT calculations of these sampled atomic configurations to the relaxation trajectories, another MLIP was parameterized (denoted as Sample-CHGNet) to perform MD simulations for enhanced sampling.

\textit{MD trajectories} -- To further sample the thermally excited atomic configurations at finite temperatures, we employed Sample-CHGNet to perform molecular dynamics (MD) simulations on disordered structures across various compositions. These structures were first generated in the pristine (lithiated) state from the CE-MC simulations at $T = 1273$ K using supercells of $10\times$ and $20\times$ formula units (f.u.), including Li$_{1.2}$Mn$_{0.5}$Ti$_{0.3}$O$_{1.9}$F$_{0.1}$, Li$_{1.2}$Mn$_{0.6}$Ti$_{0.2}$O$_{1.8}$F$_{0.2}$, Li$_{1.1}$Mn$_{0.8}$Ti$_{0.1}$O$_{1.9}$F$_{0.1}$, and Li$_{1.1}$Mn$_{0.7}$Ti$_{0.2}$O$_{2.0}$. For each disordered structure, three different delithiated levels (removing 0.2/0.4/0.6 Li per f.u. from the pristine structure) were generated to initiate the MD simulations using the Sample-CHGNet. The MD simulations were run at $T = 1000/1500/2000$ K for $t=1$ ns with a time step of $\Delta t = 2$ fs under NVT ensembles. From each MD trajectory, 200 structures were randomly sampled, and their energies were calculated using static DFT and added to the training set. In addition to the training set, we further ran CHGNet-MD simulations in NVT ensembles at $T=1000$ K using Sample-CHGNet with a supercell of Li$_{44-x}$Mn$_{32}$Ti$_{4}$O$_{76}$F$_{4}$ ($40\times$ f.u.). These trajectories were further evaluated by r$^2$SCAN-DFT static calculations and were set as the test set. 

The final CHGNet for production (denoted as Product-CHGNet) was fine-tuned using all structures in the training set (including 88,852 structures) starting from the pretrained model. It achieved a convergence of 2 meV/atom in energy, 68 meV/\AA\ in force, 0.086 GPa in stress, and 0.019 $\mu_B$ in magnetic moment for the test error. A summary flowchart is illustrated in Fig.~\ref{fig:finetune_LMTOF} to describe the whole procedure.

\begin{figure*}[t]
\centering
\includegraphics[width=\linewidth]{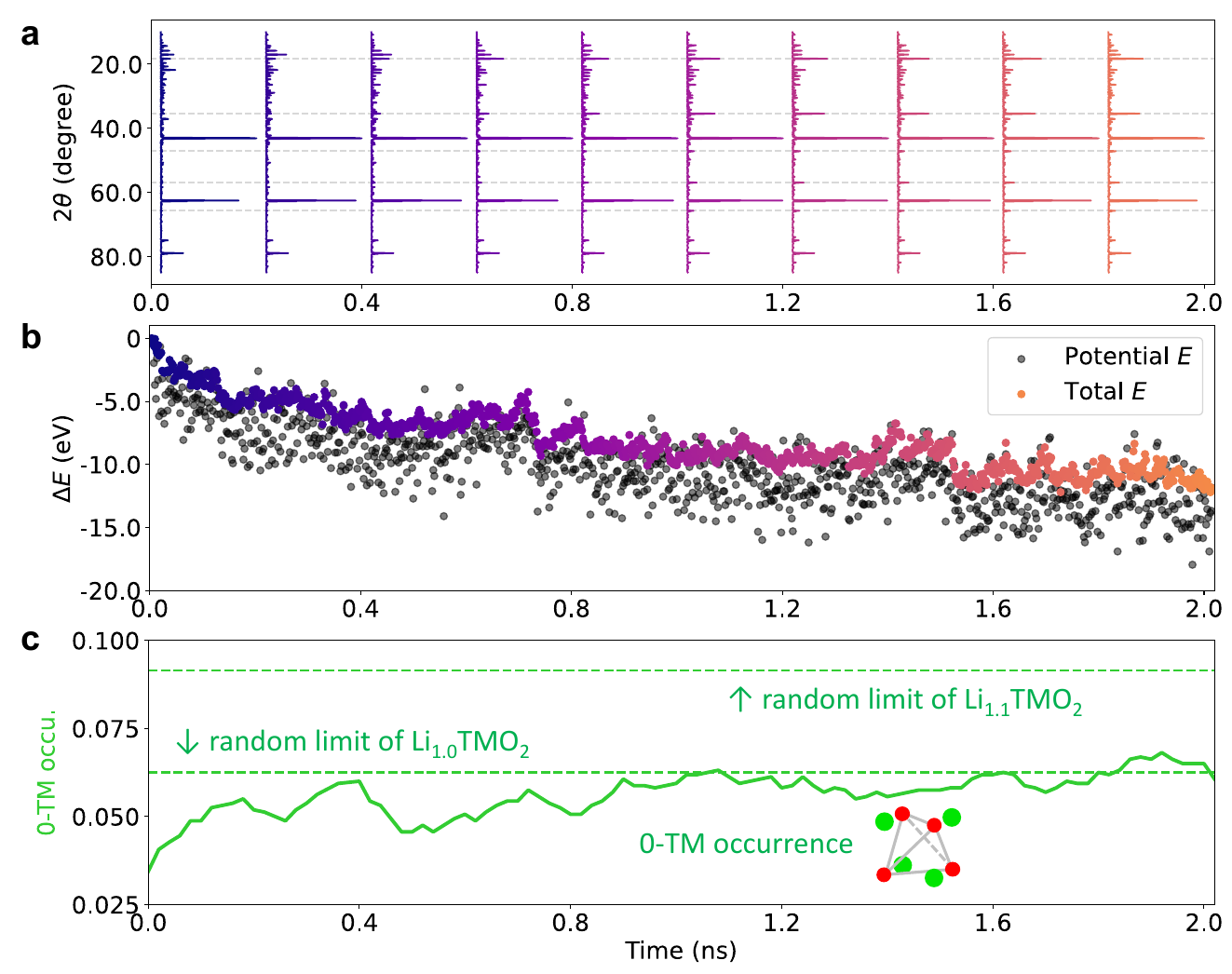}
\caption{\textbf{Structural and energetic analysis from the MD simulation}. (a) Simulated X-ray diffraction (XRD) patterns of the refined structures, with gray dashed lines indicating characteristic spinel ordering peaks (18$^\circ$, 35$^\circ$, etc.). (b) Energy evolution showing potential energy (black dots) and total energy (colored dots) relative to the initial structure, normalized per anion. (c) Evolution of 0-TM tetrahedra per formula unit, with green dashed lines indicating reference 0-TM tetrahedra occurrence in random Li$_{1.0}$TMO$_2$ and Li$_{1.1}$TMO$_2$ structures.}
\label{fig:MD_traj_stat} 
\end{figure*}

\begin{figure*}[t]
\centering
\includegraphics[width=\linewidth]{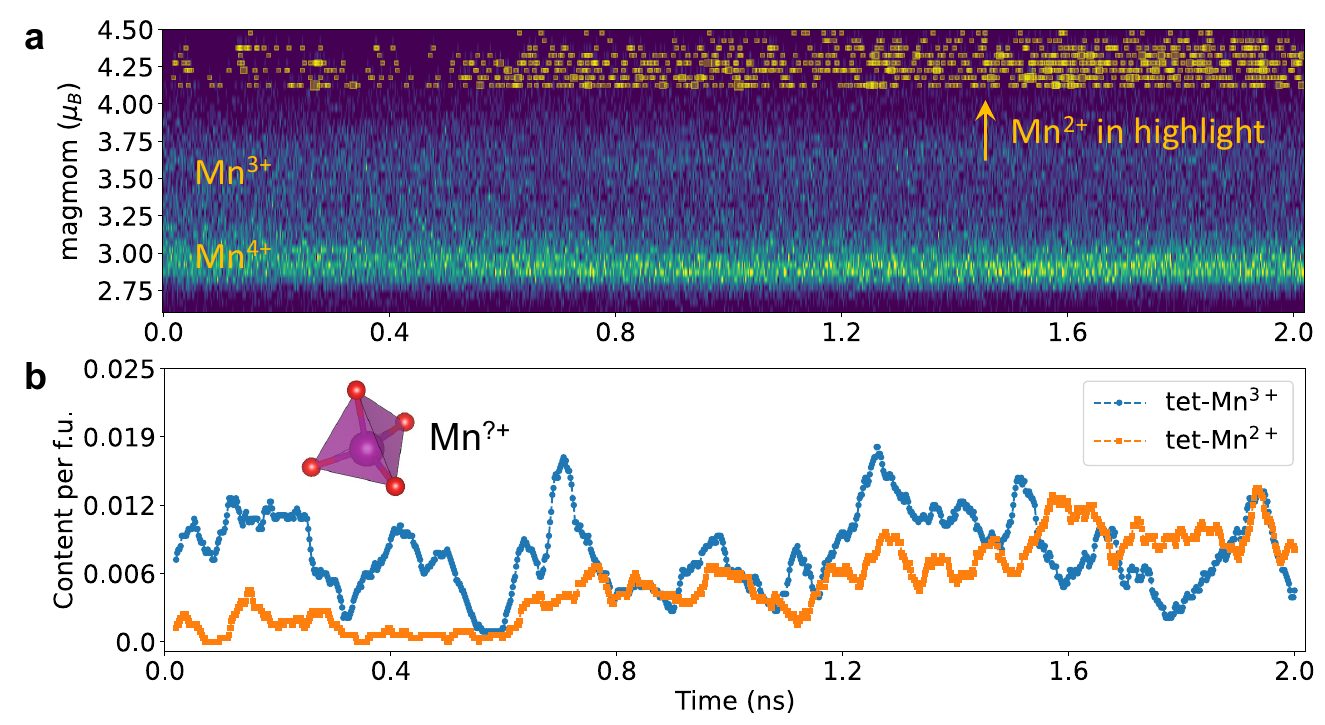}
\caption{\textbf{Atomic charge analysis of Mn from the MD simulation and magnetic moment prediction}. (a) Time-resolved histogram of magnetic moments for all Mn ions, with brighter colors indicating higher Mn ion populations at given magnetic moment values. Yellow circles highlight the Mn$^{2+}$ distribution. Valence states are categorized as Mn$^{2+} \in [4.1\mu_B, 5.0\mu_B]$, Mn$^{3+} \in [3.25\mu_B, 4.1\mu_B]$, and Mn$^{4+} \in [2.5\mu_B, 3.25\mu_B]$. (b) Temporal evolution of charge-decorated Mn occupancy on tetrahedral sites, showing Mn$^{3+}$ (blue line) and Mn$^{2+}$ (orange line) content per formula unit. The increase in tetrahedral Mn$^{2+}$ after $\sim0.6$ ns coincides with spinel-like ordering formation.}
\label{fig:MD_traj_chg} 
\end{figure*}

\section{Results}

To investigate the structural evolution as a disordered rocksalt transforms, we performed charge-informed molecular dynamics from the disordered structure to monitor the phase change to the $\delta$-phase. The initial (disordered) structure was generated by sampling from the equilibrium ensemble of CE-MC simulations at $T = 1273$ K. The MC structure represents the pristine (lithiated) composition (Li$_{1.1}$Mn$_{0.8}$Ti$_{0.1}$O$_{1.9}$F$_{0.1}$) with a supercell containing 320 atoms. Based on the pristine structure, a delithiated structure (Li$_{0.5}\square_{0.6}$Mn$_{0.8}$Ti$_{0.1}$O$_{1.9}$F$_{0.1}$, where $\square$ represents a cation vacancy) was then obtained by randomly removing Li ions and relaxing the delithiated configuration using the Product-CHGNet with a force convergence criterion of $<0.1$ eV/\AA. After relaxation, MD simulations were subsequently performed for $t=2$ ns with a time step of $\Delta t = 2$ fs under the NVT ensemble at $T = 1273$ K. The MD trajectories were collected for subsequent structural analysis.

\subsection{Structure ordering analysis}

Figure~\ref{fig:MD_traj_stat}a presents the simulated X-ray diffraction (XRD) patterns derived from the structures within the time interval of $t=0$ to $t=2$ ns. 
To analyze the evolution of cation ordering on the FCC lattice, the atoms in the MD snapshots were moved to their nearest ideal Wyckoff sites.
The gray dashed lines in Fig.~\ref{fig:MD_traj_stat}a denote the characteristic peaks of spinel ordering (18$^\circ$, 35$^\circ$, etc.). At the onset of the MD simulation ($t=0$ ns), there are no obvious peaks indicating spinel ordering. After approximately $t=0.6$ ns, the intensity of the spinel peaks increases and becomes more significant as the MD simulation proceeds, suggesting the formation of the spinel-like $\delta$-phase.

Figure \ref{fig:MD_traj_stat}b illustrates the energy landscape across the simulated MD trajectories. The black points denote the potential energy, while the colored points represent the total energy. The energies are depicted relative to their values at the initial state. A notable decrease in energy (approximately 60 meV/anion) is observed after $t=2$~ns when the $\delta$-phase has formed. 
This energy reduction indicates that spinel-like cation ordering is thermodynamically favored at the composition Li$_{0.6}$Mn$_{0.8}$Ti$_{0.1}$O$_{1.9}$F$_{0.1}$ when sufficient cation vacancies are present.
This thermodynamic preference for spinel-like ordering aligns with previous computational studies that demonstrated the stability of spinel structures in Li-Mn-O systems, such as Li$_{0.5}$MnO$_2$ \cite{wang2007first}.

To study the change in local cation ordering, we analyzed the short-range order (SRO) indicated by the tetrahedron occupancy as a function of MD simulation time. Figure \ref{fig:MD_traj_stat}c presents the occurrence of 0-TM (solid green line) tetrahedra in the MD structure, as a ratio to all tetrahedral sites. 
To compare with disordered phases, the dashed green lines in Fig.~\ref{fig:MD_traj_stat}c indicate the random limit of the 0-TM tetrahedra for two Li contents ($\sim 0.06$ for Li$_{1.0}$ and $\sim 0.09$ for Li$_{1.1}$ at the fully lithiated level) for reference. At the beginning of the MD simulation ($t=0$ ns), the occurrence of 0-TM tetrahedra is approximately 0.03. This value is significantly lower than that of the random limit for Li$_{1.1}$TMO$_2$ (0.09). The low value of 0-TM content in the initial structure indicates that SRO is unfavorable for long-range Li percolation, which is consistent with previous predictions in Mn-Ti-based DRXs \cite{Ji2019_hidden_SRO, Ouyang2020_SRO}. As the phase transformation progresses, the occurrence of 0-TM tetrahedra increases, reaching approximately 0.06 after $t=2$ ns of MD simulation. This rise in the number of 0-TM tetrahedra agrees with the high 0-TM content in the spinel structure, indicating the emergence of spinel-like ordering.

\subsection{Charge distribution with ion rearrangement}

To understand the correlation between Mn migration and its valence state \cite{Cai2023_NatEnergy, Reed_Ceder_Ven_2001}, we inferred atomic charge states from the magnetic moments predicted by CHGNet.
Figure~\ref{fig:MD_traj_chg}a presents the site-wise distribution of magnetic moments for Mn ions. The valence states of Mn are categorized as follows: Mn$^{2+} \in [4.1\mu_B, 5.0\mu_B]$, Mn$^{3+} \in [3.25\mu_B, 4.1\mu_B]$, and Mn$^{4+} \in [2.5\mu_B, 3.25\mu_B]$. At the beginning of the MD simulation, Mn predominantly exists in valence states of +3 and +4, consistent with the partial delithiation of the material. As the phase transformation progresses, the concentration of Mn$^{2+}$ gradually increases, as shown by the increased frequency of the highlighted circles in Fig.~\ref{fig:MD_traj_chg}a, particularly after the emergence of spinel-like ordering from 0.6 to 2.0 ns.

For a detailed description of Mn migration conditioned on its valence state, the occupancy of Mn$^{3+}_{\text{tet}}$ and Mn$^{2+}_{\text{tet}}$ per f.u. is plotted in Fig.~\ref{fig:MD_traj_chg}b. Although previous NEB calculations reveal that Mn$^{2+}$ has the lower migration barrier of all Mn valence states \cite{Reed2001_LiMnO, Cai2023_NatEnergy}, the atomic charges predicted by CHGNet show that most of the tetrahedral Mn are Mn$^{3+}_{\text{tet}}$ from $t=0$ to $0.6$ ns during the MD simulation. This suggests that Mn$^{3+}$ may migrate without the need for charge disproportionation ($2\text{Mn}^{3+} \rightarrow \text{Mn}^{2+} + \text{Mn}^{4+}$) as proposed by \citet{Reed2001_LiMnO}. The rise in Mn$^{2+}_{\text{tet}}$ appears after around $t\approx0.6$~ns, which corresponds to the formation of spinel characteristic peaks (18$^\circ$, 35$^\circ$) in Fig.~\ref{fig:MD_traj_stat}a. 
This evidence suggests that Mn$^{3+}$ migration initiates the transformation before long-range spinel-like order is established. The subsequent emergence of tetrahedral Mn$^{2+}$ appears to be a consequence of this long-range ordering rather than a prerequisite for local cation hops.
The simulation results are consistent with previous MD simulations of the Li$_x$MnO$_2$ phase transformation revealed by \citet{Deng2023_chgnet} using the GGA$+U$-based MLIP.

\subsection{Intercalation voltage profile of $\delta$-phase}

To evaluate the effect of partial disorder on electrochemical performance, we computed the intercalation voltage profiles of the transformed $\delta$-phase structure using Product-CHGNet. Specifically, we focus on the voltage characteristic of ordered spinel Li$_x$Mn$_2$O$_4$ at lower voltage (for $x \geq 1$), in which the occupancies of Li ions are transformed collectively from the $8a$ tetrahedral sites to the empty $16c$ octahedral sites upon lithiation. The reason this happens collectively in a well-ordered spinel is that $8a$ (tet) and $16c$ (oct) sites cannot be simultaneously occupied. Hence, upon lithiation the system empties the $8a$ sites and occupies $16c$. However, in a (partially) disordered rocksalt structure, the $8a$, $16c$, and $16d$ crystallographic labels cannot be defined precisely.

For this reason, we define a ``0-TM" site as any interstitial tetrahedral site that is exclusively face-shared by four non-transition metal cations (i.e., Li or vacancy). This definition captures the essential local environment of the $8a$ site in an ordered spinel which also has no transition metal neighbors. It also generalizes to disordered rocksalts where the connectivity of 0-TM sites is essential for evaluating the percolating Li content \cite{Urban2014_AEM_order}. Using this more general description, we can expect two key intercalation processes:

(a)~A lower-voltage reaction whereby a tetrahedral Li is removed and its octahedral sites become filled with Li. For such an isolated 0-TM site, this reaction can be written as:
\begin{equation}
3 \square_{\text{oct}} + 1 \text{Li}_{\text{tet}} \leftrightarrow 4 \text{Li}_{\text{oct}} ~ (\text{0-TM}) 
\label{eq:0TM_tetLi_conversion}
\end{equation}

(b)~A higher-voltage reaction corresponding to the direct insertion of Li into an empty 0-TM tetrahedral site:
\begin{equation}
    \square_{\text{tet}} \leftrightarrow \text{Li}_{\text{tet}}
\label{eq:tetLi_direct}
\end{equation}
This framework allows us to consistently compare the electrochemical behavior of the ordered spinel, disordered rocksalt, and the intermediate $\delta$-phase.

\begin{figure}[htb]
\centering
\includegraphics[width=\linewidth]{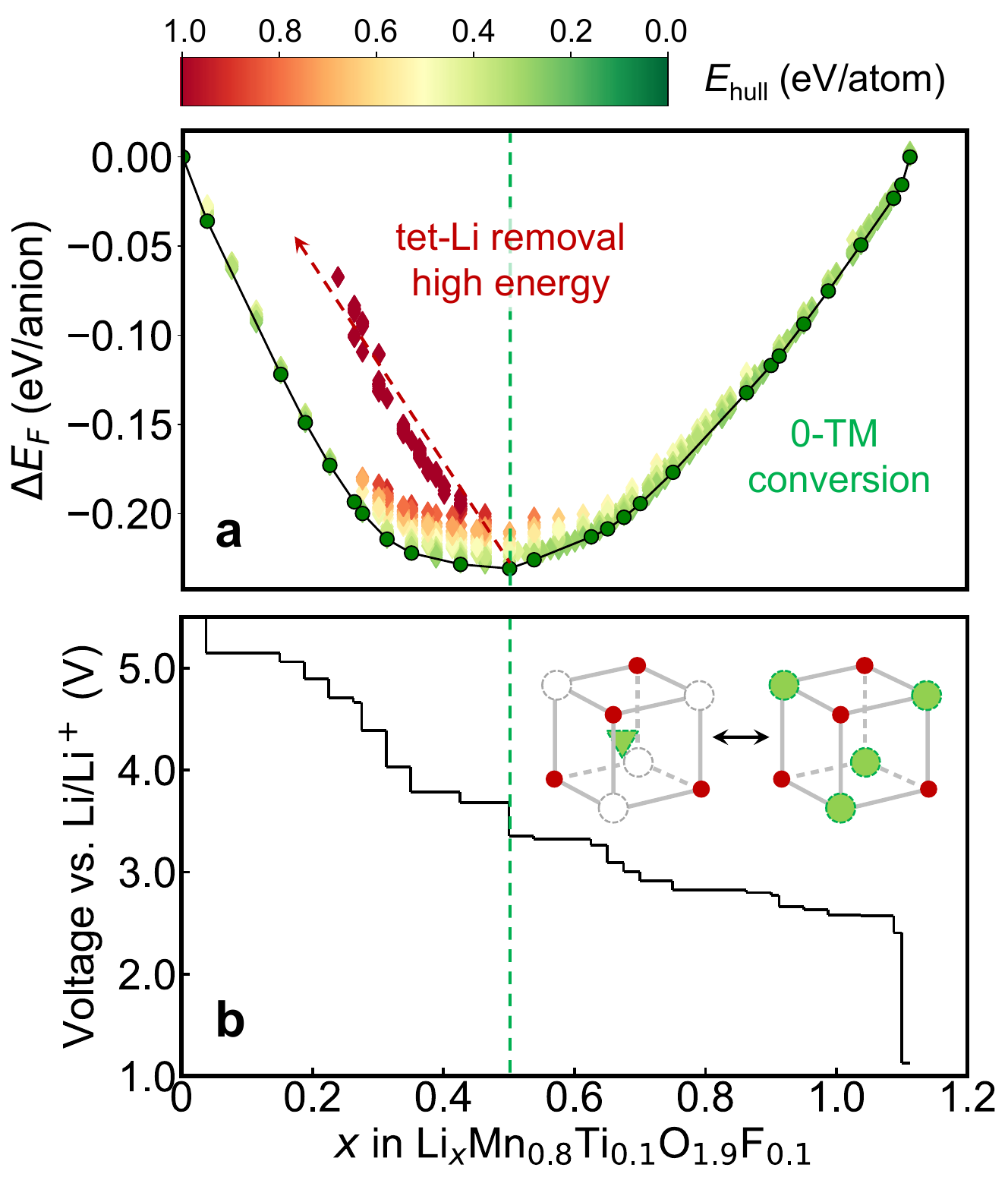}
\caption{\textbf{Phase diagram and intercalation voltage profiles of LMTOF in the $\delta$-phase}. (a) Formation energy of various Li-vacancy configurations as a function of Li concentration, with color indicating energy above the convex hull (black line). The color bar indicates the energy above the convex hull. Ground states on the convex hull are marked with green circles. The dark red arrow indicates high formation energy ($\Delta E_F$) for direct removal of Li$_{\text{tet}}$ when $x \leq 0.5$. (b) Voltage profiles derived from the convex hull. The inset shows the $4 \text{Li}_{\text{oct}}$ (0-TM) $\rightarrow \text{Li}_{\text{tet}}$ conversion reaction scheme used to evaluate high Li content intercalation (green spheres = Li cations, red spheres = O/F anions).}
\label{fig:delta_V_profile} 
\end{figure}

We generated a series of Li-vacancy configurations to simulate the electrochemical (de)lithiation process topotactically. Starting with the fully lithiated structure obtained from MD, we held the TM and anion sublattices fixed. At each (de)lithiation level, distinct Li-vacancy orderings were created by selecting different 0-TM sites within the host structure and applying the local transformation described in Eq.~\eqref{eq:0TM_tetLi_conversion} and \eqref{eq:tetLi_direct}. The Product-CHGNet was used to perform structure relaxation for both lithiation and delithiation paths (see Appendix \ref{appendix:Li-vac}). The energies of these (de)lithiated structures were used to construct the convex hull and the corresponding voltage profile as shown in Fig.~\ref{fig:delta_V_profile}a and Fig.~\ref{fig:delta_V_profile}b.

\begin{figure*}[t]
\centering
\includegraphics[width=\linewidth]{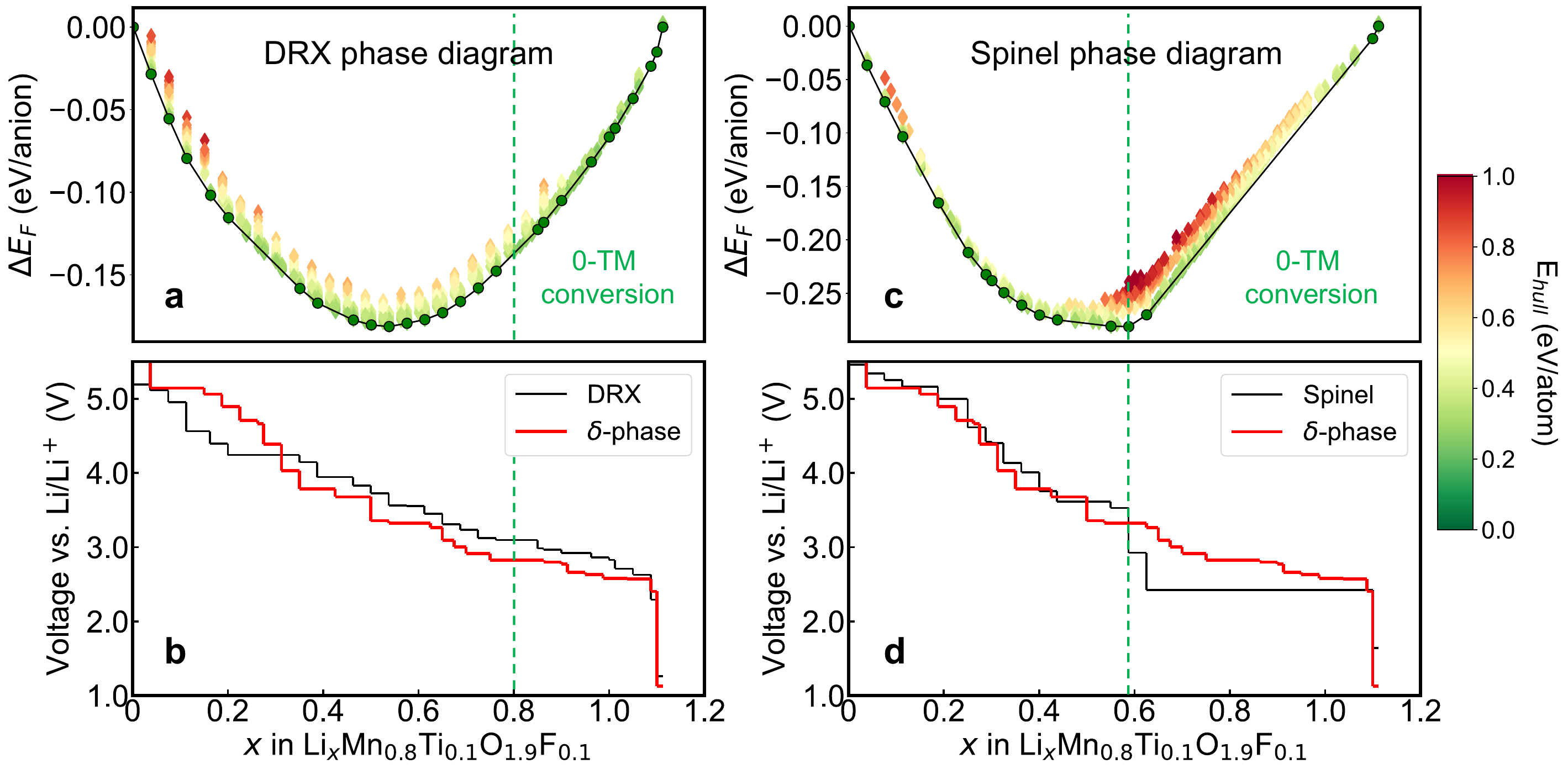}
\caption{\textbf{Phase diagram and intercalation voltage profiles of LMTOF in DRX and spinel structures}. (a) Formation energy of various Li-vacancy configurations as a function of Li concentration, with color indicating energy above the convex hull (black line). (b) DRX LMTOF voltage profile constructed from convex hull ground states, compared with $\delta$-phase profile (red line). Green dashed line indicates the limit of the Li capacity from the $4 \text{Li}_{\text{oct}}$ (0-TM) $\leftrightarrow\text{Li}_{\text{tet}}$ conversion reaction. (c) Formation energy convex hull of the spinel structure using the identical color scheme as (a). (d) Spinel voltage profile showing two-phase reaction between Li$_{1.1}$ and Li$_{0.6}$ compositions, evidenced by the linear convex hull in (c) and flat voltage plateau in (d). The spinel LMTOF exhibits a $>1$ V voltage step near Li$_{0.6}$, while the $\delta$-phase (red line) shows solid-solution behavior.}
\label{fig:drx_spinel_V_profile}
\end{figure*}

For the region with high Li concentration ($0.5 \leq x \leq 1.1$), the 0-TM to $\text{Li}_{\text{tet}}$ conversion exhibits nearly solid-solution behavior, as evidenced by the convex hull being connected by several ground states in Fig.~\ref{fig:delta_V_profile}a. Particularly, for the early stage of delithiation ($0.7 < x < 1.1$), the voltage exhibits a pseudo plateau with minor slope, indicating that the delithiation through Eq. \eqref{eq:0TM_tetLi_conversion} occurs in similar local chemical environments and exhibits similar chemical potentials (voltages). The computed voltage at $x = 1.1$ is around 2.40 V, which is slightly lower than the experimental value. 
This discrepancy is likely attributable to the residual self-interaction error common in DFT functionals, even for r$^2$SCAN, when a Hubbard U correction is not applied \cite{Zhou2004_LDAU, Swathilakshmi2023_r2scan_U}.
Although the SCAN-based functional removes more self-interaction than previous LDA and GGA functionals \cite{Sun2015_SCAN}, it does not fully remove it and underestimates the intercalation voltage, which is consistent with previous reports \cite{Chen2023_remove, Zhong2023_PRXEnergy}.

Upon reaching $x = 0.5$, lithium occupies all tetrahedral 0-TM sites (Li$_{\text{tet}}$) in addition to the octahedral Li$_{\text{oct}}$ not in the 0-TM sites. Two competing delithiation processes are available, involving either the removal of Li$_{\text{tet}}$ or Li$_{\text{oct}}$.  We computed both pathways and found that the low-energy configurations near the convex hull in the range $0.25 \leq x \leq 0.5$ correspond to the removal of Li$_{\text{oct}}$, while the high-energy configurations (represented by dark red markers in Fig.~\ref{fig:delta_V_profile}a) correspond to the removal of Li$_{\text{tet}}$. These results indicate that even in a structure with partial spinel ordering, Li$_{\text{oct}}$ has a preference for delithiation before Li$_{\text{tet}}$. 
In contrast, the delithiation in ordered spinel Li$_x$MnO$_2$ for $0 \leq x \leq 0.5$ entirely corresponds to the removal/insertion of Li$_{\text{tet}}$ with a $\sim 4$ V plateau. We also observe a voltage increase of $\sim 0.32$ V around $x=0.5$ (the end of 0-TM to Li$_{\text{tet}}$ conversion). The magnitude is not as large as that in spinel LiMn$_2$O$_4$ ($> 1.0$ V) or in the spinel-ordered Li$_{x}$Mn$_{0.8}$Ti$_{0.1}$O$_{1.9}$F$_{0.1}$ (see Fig.~\ref{fig:drx_spinel_V_profile}d), suggesting the absence of two-phase reactions in the $\delta$-phase. 

To further demonstrate the effect of partial disorder, we evaluated the intercalation voltage profiles of Li$_{1.1}$Mn$_{0.8}$Ti$_{0.1}$O$_{1.9}$F$_{0.1}$ in both the DRX and spinel phases as a comparison. The DRX structure was selected from the CE-MC simulations (i.e., the configuration at $t = 0$ s in the MD simulation). The spinel Li$_{1.1}$Mn$_{0.8}$Ti$_{0.1}$O$_{1.9}$F$_{0.1}$ was generated through a separate CE-MC simulation without exchanges between the $16c$ and $16d$ sites, so that all $16c$ sites are occupied by Li$_{1.0}$ and $16d$ sites are occupied by Mn$_{0.8}$, Ti$_{0.1}$, and Li$_{0.1}$, similar to a Li-excess lithiated spinel. The MC simulation allows site exchange within $16d$ sites and anion sites, which is designed to capture the underlying SRO of anions and cations in DRXs (e.g., Li-F SRO) \cite{Zhong2020_Mg}. The intercalation at high Li content in both DRX and spinel phases only included the 0-TM site conversion reaction as described in Eq.~\eqref{eq:0TM_tetLi_conversion}.

Figure~\ref{fig:drx_spinel_V_profile}a and \ref{fig:drx_spinel_V_profile}b illustrate the formation energy convex hull and the voltage profile of the DRX phase, with the $\delta$-phase plotted as a red line for comparison. As there are few 0-TM sites in the DRX structure, a relatively limited Li capacity from the 0-TM conversion is achieved (approximately 0.3 per f.u., as denoted by the vertical green dashed line in Fig.~\ref{fig:drx_spinel_V_profile}a and \ref{fig:drx_spinel_V_profile}b. The absence of spinel-like ordering results in a profile marked by a mildly steep voltage increase as delithiation progresses. Notably, the overall voltage profile of the DRX phase between $0.3 < x < 1.1$ is higher than that of the $\delta$-phase, suggesting a restricted amount of accessible capacity around $\sim 3$ V \cite{Cai2023_NatEnergy, Ahn2023_AEM}. In contrast, Figure \ref{fig:drx_spinel_V_profile}d depicts the calculated results for spinel Li$_{x}$Mn$_{0.8}$Ti$_{0.1}$O$_{1.9}$F$_{0.1}$. The voltage profile clearly displays a two-phase reaction between Li$_{1.1}$ and Li$_{0.6}$ consistent with the linear convex hull in Fig.~\ref{fig:drx_spinel_V_profile}c. Due to the absence of disorder, the spinel phase exhibits a lower average voltage with a two-phase reaction between $0.6 < x < 1.1$, and a voltage step larger than 1 V at the end of 0-TM conversion. For $x < 0.4$, the intercalation corresponds to  Li$_{\text{tet}} \rightarrow \square_{\text{tet}}$, which shows a similar high-voltage feature as the $\delta$-phase. The effect of cation disorder on the spinel structure demonstrated in our simulations is in agreement with the previous findings on the cation disorder effects in LiMnO$_2$ spinel by \citet{Chen2023_remove}.

\section{Discussion}

In the evolving landscape of modern battery materials, the multirole of components and complexity of configurations exemplified by DRX cathode materials necessitates advanced computational models. Particularly, Mn-rich DRXs exhibit a phase transformation to a partially disordered spinel phase ($\delta$-phase) via in situ electrochemical cycling \cite{Cai2023_NatEnergy, Ahn2023_AEM, Holstun2025_pulse} or chemical delithiation \cite{Hau2024_delta, Li2024_structural_Evolution_DRX}. 
This transformation is challenging for conventional modeling approaches, as it requires capturing the kinetic pathway of TM migration in the disordered environment. While \textit{ab initio} molecular dynamics could theoretically address this issue, its $\mathcal{O}(N^3)$ computational cost renders it impractical for probing large systems or long timescales. Alternatively, kinetic Monte Carlo (kMC) methods with CE models have been proposed to model cation ordering changes in both cathodes and solid electrolytes \cite{Xiao2019_kmc, Deng2023_kmc}. However, parameterizing CEs for disordered oxides/oxyfluorides and accurately calculating migration barriers in diverse local environments with DFT remains a non-trivial endeavor \cite{Yang2022_npj}.

To address these challenges, we fine-tuned a foundation MLIP with sampled r$^2$SCAN-DFT calculated energies, forces, stresses, and magnetic moments. The multi-step PES sampling captures the complex interactions from diverse chemical environments and different valence states of TMs.
Our MD simulations using the fine-tuned CHGNet successfully captured the phase transition from a disordered to a spinel-like configuration in Li$_{0.6}$Mn$_{0.8}$Ti$_{0.1}$O$_{1.9}$F$_{0.1}$.

The simulations reveal that the formation of the $\delta$-phase from DRX promotes the increase of 0-TM configurations on tetrahedra, which play a crucial role in facilitating long-range Li-ion transport. This result is consistent with the improved high-rate performance observed in experiments \cite{Cai2023_NatEnergy, Ahn2023_AEM, Hau2024_delta}. By analyzing the simulated electrochemical intercalation voltage profiles of the $\delta$-phase and comparing them to those of the DRX and spinel phases, two conclusions can be drawn: (1) The presence of partial cation disorder in the $\delta$-phase effectively eliminates the two-phase reaction, as evidenced by the absence of a sharp voltage increase after $\sim$ Li$_{0.5}$ removal (see Fig.~\ref{fig:drx_spinel_V_profile}d). This is consistent with earlier predictions on partially disordered spinels \cite{Chen2023_remove}. (2) The introduction of spinel-like ordering increases the 0-TM occurrence (as indicated by the increased number of 0-TM tetrahedra in Fig.~\ref{fig:MD_traj_stat}c) and thus expands the available Li capacity from the 0-TM to Li$_{\text{tet}}$ conversion reaction around 3 V (denoted by the green dashed line in Fig.~\ref{fig:delta_V_profile}b). The Li capacity from 0-TM to Li$_{\text{tet}}$ conversion in the $\delta$-phase is even higher than that of the spinel in Fig.~\ref{fig:drx_spinel_V_profile}d since the 0-TM sites in partially disordered phases are less spatially correlated with each other compared to the spinel, where each Li$_{\text{oct}}$ is shared by two 0-TM sites.

Although this study represents an integrated investigation of phase transformation in complex oxides combining MC simulations for configurational thermodynamics and MLIP-MD simulations for kinetic modeling, several limitations were identified that could be further addressed in future studies. 
For example, while our MLIP enables nanosecond-scale or million-atoms simulations with \textit{ab initio} accuracy \cite{Han_2025_DistMLIP}, many important physical processes—such as the formation of anti-phase boundaries in the $\delta$-phase \cite{Hau2024_delta} or phase segregation in Li$_x$FePO$_4$ cathodes \cite{Zhao2023_LFP}---occur on timescales still beyond this reach.
Achieving \textit{ab initio} accuracy while enabling simulations beyond typical MD timescales (such as using generative models for dynamic trajectories) holds promise for investigating these processes \cite{fu2022simulate, Hsu2024_score, nam2024flow}. Regarding model fine-tuning, recent advances such as active learning with Bayesian force fields \cite{Xie2023_flare}, stratified sampling from MLIP-driven MD \cite{qi2024_robust}, and diffusion-based generative models for mapping reaction pathways \cite{duan2023accurate, Zhao2025_harness} provide more efficient sampling techniques beyond the passive sampling implemented in this work, which could enhance MLIP training accuracy by exploring the PES more effectively. 
Finally, it is important to clarify that the term "charge" refers to the valence state (nominal charge) of the TM, which is inferred from the predicted magnetic moments. The current model does not explicitly compute long-range electrostatic (Coulombic) interactions. Looking forward, the model's capabilities could be significantly enhanced by incorporating response charge prediction, such as Latent Ewald Summation (LES) \cite{Cheng2025_LES}. LES integrates long-range electrostatics and the prediction of Born effective charges \cite{zhong_machine_2025}, which has demonstrated a universal augmentation for mainstream MLIPs \cite{kim_universal_2025}. We believe an LES-augmented CHGNet would yield a more physically robust description of charge, thereby improving the modeling of complex phenomena such as cation migration during phase transformations in oxides.

In summary, we demonstrated the application of a MLIP to model the complex phase transformation from DRX to spinel-like $\delta$-phase in Li-Mn-Ti-O-F systems. The MLIP-MD simulations revealed the mechanism of spinel-like ordering formation and its effect on electrochemical properties. As computational capabilities and new methodologies emerge--particularly in accelerating MLIP simulations and in generative modeling--these advances will pave the way for more accurate and efficient modeling of phase transformations in new energy storage materials.

\section{Acknowledgments}
This work was primarily supported by the U.S. Department of Energy, Office of Science, Office of Basic Energy Sciences, Materials Sciences and Engineering Division under Contract No. DE-AC0205CH11231 (Materials Project program KC23MP), and the Assistant Secretary for Energy Efficiency and Renewable Energy, Vehicle Technologies Office, under the Advanced Battery Materials Research (BMR) Program of the US Department of Energy (DOE) under contract No. DE-AC0205CH11231. The computational modeling in this work was supported by the computational resources provided by the Extreme Science and Engineering Discovery Environment (XSEDE), supported by National Science Foundation grant number ACI1053575; the National Energy Research Scientific Computing Center (NERSC); the National Renewable Energy Laboratory (NREL) clusters under the drx allocation, and the Lawrencium computational cluster resource provided by the IT Division at the Lawrence Berkeley National Laboratory. The authors thank Zijian Cai for the valuable discussions.

\section*{Data Availability}

The supporting data and code are available at \url{https://github.com/zhongpc/drx2delta}.

\appendix
\numberwithin{equation}{section}
\section{DFT Calculations}\label{appendix:DFT}

All DFT calculations were performed with the \texttt{VASP} package using the projector-augmented wave method \cite{Kresse1996_VASP, Kresse1999_PAW}, a plane-wave basis set with an energy cutoff of 680 eV, and a \textit{k}-point sampling grid spacing of 0.25 \AA$^{-1}$. The calculations were converged to $10^{-6}$ eV in total energy for electronic loops and 0.02 eV/\AA\ in interatomic forces for ionic loops. The regularized strongly constrained and appropriately normed meta-GGA exchange-correlation functional (r$^2$SCAN) \cite{Sun2015_SCAN, Furness2020_r2SCAN} was used with consistent computational settings as \texttt{MPScanRelaxSet} \cite{Kingsbury2022_r2scan_PRM}. 

\section{Cluster Expansion \& Monte Carlo}\label{appendix:CE}

The cluster expansion (CE) method \cite{Sanchez1984} is used to study the configurational thermodynamics of materials in which sites can be occupied by multiple cations, and has been applied to study the Li-vacancy intercalation chemistry in layered and disordered materials \cite{Wolverton1998_LiCoO2, Zhong2023_PRXEnergy}. The CE expands the energy as a sum of many-body configurational interactions:
\begin{equation}
	E(\boldsymbol{\boldsymbol{\sigma}}) = \sum_{\beta} m_{\beta} J_{\beta}\left\langle \Phi_{\boldsymbol{\alpha} \in \beta} \right\rangle_{ \beta} + \frac{E_0}{\varepsilon_r},~ \Phi_{\boldsymbol{\alpha}} = \prod_{i=1}^N \phi_{\alpha_i}(\sigma_i)
	\label{eq:ce_definition}
 \end{equation}
A configuration $\boldsymbol{\sigma}$ represents a specific occupancy state of species its allowed sites, and $\sigma_i$ describes which species sit on the $i$-th site of the lattice.  The cluster basis function $\Phi_{\boldsymbol{\alpha}} = \prod_{i=1}^N \phi_{\alpha_i}(\sigma_i)$ is the product of site basis functions $\phi_{\alpha_i}(\sigma_i)$ across a collection $\alpha$ of multiple sites \cite{VandeWalle2009}. The average is taken over the crystal symmetry orbits $\beta$, forming a complete basis to expand the energy function in the configuration space. The expansion coefficients $J_{\beta}$ are called effective cluster interactions (ECIs). The electrostatic energy (Ewald energy $E_0/\varepsilon_r$) is included to capture long-range electrostatic interactions ($E_0$ is the unscreened electrostatic energy, and $1/\varepsilon_r$ is fitted as one of the ECIs).

For the simulation of atomic orderings, a cluster-expansion Hamiltonian was generated in the chemical space of Li$^+$-Mn$^{3+}$-Ti$^{4+}$- O$^{2-}$-F$^-$, with pair interactions up to 7.1 \AA, triplet interactions up to 4.0 \AA, and quadruplet interactions up to 4.0 \AA\ based on a primitive cell of the rocksalt structure with lattice parameter $a=3$ \AA.  In total, 83 ECIs (including the constant term $J_0$) were defined, and the CE Hamiltonian was fitted with 322 unique structures. The ECIs were determined with the optimal sparseness and prediction error (with 69 non-zero elements and $<$ 8 meV/atom) with an $\ell_0\ell_2$-norm regularized regression \cite{Zhong2022L0L2}. We refer readers to Ref. \cite{Barroso-Luque2022_theory} for details of CE in ionic systems. The CE constructions were performed using \texttt{smol} \cite{Barroso-Luque2022smol} and \texttt{pymatgen} for structure processing \cite{Ong2013_pymatgen}. Metropolis--Hastings algorithm was used to sample the atomic configurations from the equilibrium canonical ensemble in CE-MC simulations.

\section{Li-vacancy phase diagram}\label{appendix:Li-vac}

To construct the simplified intercalation voltage profiles, we used an enumeration method for the generation of the delithiated structure and the construction of Li-vacancy phase diagrams. For each delithiation level, the following steps are performed:

(1) Select an initial structure with composition Li$_{x_1}$TMO$_2$, relax the structure using Product-CHGNet, and obtain the internal energy $E_{\text{Li}_{x_1}\text{TM}\text{O}_2}$.

(2) Generate 20 different delithiated structures, each with a different vacancy occupancy and composition Li$_{x_2^i}$TMO$_2$ ($x_2^i$ may not be equal for all $\{i\}$ structures for the 0-TM to Li$_{\text{tet}}$ conversion). This generation process is referred to as an \emph{intercalation step}.

(3) Relax these 20 structures using Product-CHGNet and obtain their internal energies $E_{\text{Li}_{x_2^i}\text{TM}\text{O}_2}$ ($i$ denotes the index of each structure).

(4) The proposed next delithiated structure is chosen based on the minimum voltage with respect to the initial structure Li$_{x_1}$TMO$_2$ using the following equation:
\begin{equation}
\bar{V}(x_1, x_2) \approx -\frac{E_{\text{Li}_{x_1}\text{TM}\text{O}_2} - E_{\text{Li}_{x_2}\text{TM}\text{O}_2} - (x_1 - x_2)E_{\text{Li}}}{F(x_1 - x_2)},
\label{eq:avg_voltage}
\end{equation}
where $E_{\text{Li}}$ is the internal energy of the bcc Li metal, and $E_{\text{Li}_{x}\text{TM}\text{O}_2}$ represents the internal energy of the (de)lithiated structures.

(5) Use the $i$-th structure with the lowest voltage as a start and repeat Step 1 until all possible intercalation steps are consumed.

For the lithiation process, the selection criterion is changed to find the lithiated structure with the maximum voltage. It is important to note that different 0-TM sites may share the same Li atoms in an FCC framework and therefore the conversion of different 0-TM to Li$_{\text{tet}}$ thus may result in delithiated structures with varying Li concentrations in Step (2). It is more appropriate to use the minimum/maximum voltage as a selection criterion rather than the energy itself for suggesting the next step in the enumeration method.

\bibliography{references}
\end{document}